\documentclass{phb-proc4-auth}

\usepackage{graphicx}
\usepackage{amssymb}
\usepackage{amsmath}

\begin{document}

\newcommand{\green}[1]{{ \left<\!\left< {#1} \right>\!\right> }}
\newcommand{\biggreen}[1]{{ \big<\!\big< {#1} \big>\!\big> }}

\newcommand{\cdag}{c^{\dagger}}
\newcommand{\cnod}{c^{\phantom{\dagger}}}
\newcommand{\fdag}{f^{\dagger}}
\newcommand{\fnod}{f^{\phantom{\dagger}}}
\newcommand{\bdag}{b^{\dagger}}
\newcommand{\bnod}{b^{\phantom{\dagger}}}

\newcommand{\UA}{\uparrow}
\newcommand{\DA}{\downarrow}

\begin{frontmatter}


\journal{SCES '04}


\title{Phase diagram and dynamic response functions of the Holstein-Hubbard model}

%
%
%
%
%
%

\author[UK]{W. Koller}
\author[UK]{D. Meyer}
\author[UK]{A. C. Hewson}
\author[JP]{Y. \=Ono}

%
 
\address[UK]{Department of Mathematics, Imperial College, London SW7 2AZ, U.K.}
\address[JP]{Department of Physics, Niigata University, Ikarashi, Niigata 950-2181, Japan}

%
%
%
%


%
%
%
%


\begin{abstract}
We present the phase diagram and dynamical correlation functions for the
Holstein-Hubbard model at half filling and at zero temperature. The calculations
are based on the Dynamical Mean Field Theory. The effective impurity model is
solved using Exact Diagonalization and the Numerical Renormalization Group.
Excluding long-range order, we find three different paramagnetic phases,
metallic, bipolaronic and Mott insulating, depending on the Hubbard
interaction~$U$ and the electron-phonon coupling~$g$.
We present the behaviour of the one-electron spectral functions and phonon
spectra close to the metal insulator transitions.
\end{abstract}

%
%

\begin{keyword}

strongly correlated electrons, electron-phonon coupling, Mott transition

\end{keyword}


\end{frontmatter}

%
%
%
%
%

Electron-phonon effects in strongly correlated electron systems are expected
to be significant but have sofar received little theoretical attention.
We study these effects here using the Holstein--Hubbard
model\cite{KMH04pre,KMOH04} where a local Einstein phonon mode
couples linearily to local charge fluctuations.
The Hamiltonian is given by
\[                                 
  \begin{aligned}
    H =&
    \sum_{\vec{k}\sigma}
    \epsilon(\vec{k})\, \cdag_{\vec{k}\sigma} \cnod_{\vec{k}\sigma} +
    U \sum_{i} n_{i\uparrow} n_{i\downarrow} \\ &+
    \omega_0 \sum_i  \bdag_i \bnod_i +
    g\sum_i  (\bdag_i + \bnod_i) \big(n_{i\UA}+ n_{i\DA}-1 \big) \:.
    \end{aligned}
\]
We use a semi-elliptical band of width $W=4$ and focus on the particle-hole
symmetric case at zero temperature. The phonon frequency is fixed to
$\omega_0=0.2$.
We calculate the phase diagram, in the absence of long-range order, using a
number of local approximations (DMFT-NRG/ED \cite{KMOH04}, DIA \cite{Pot03}).

\begin{figure*}
    \centering
    \includegraphics[width=0.98\textwidth]{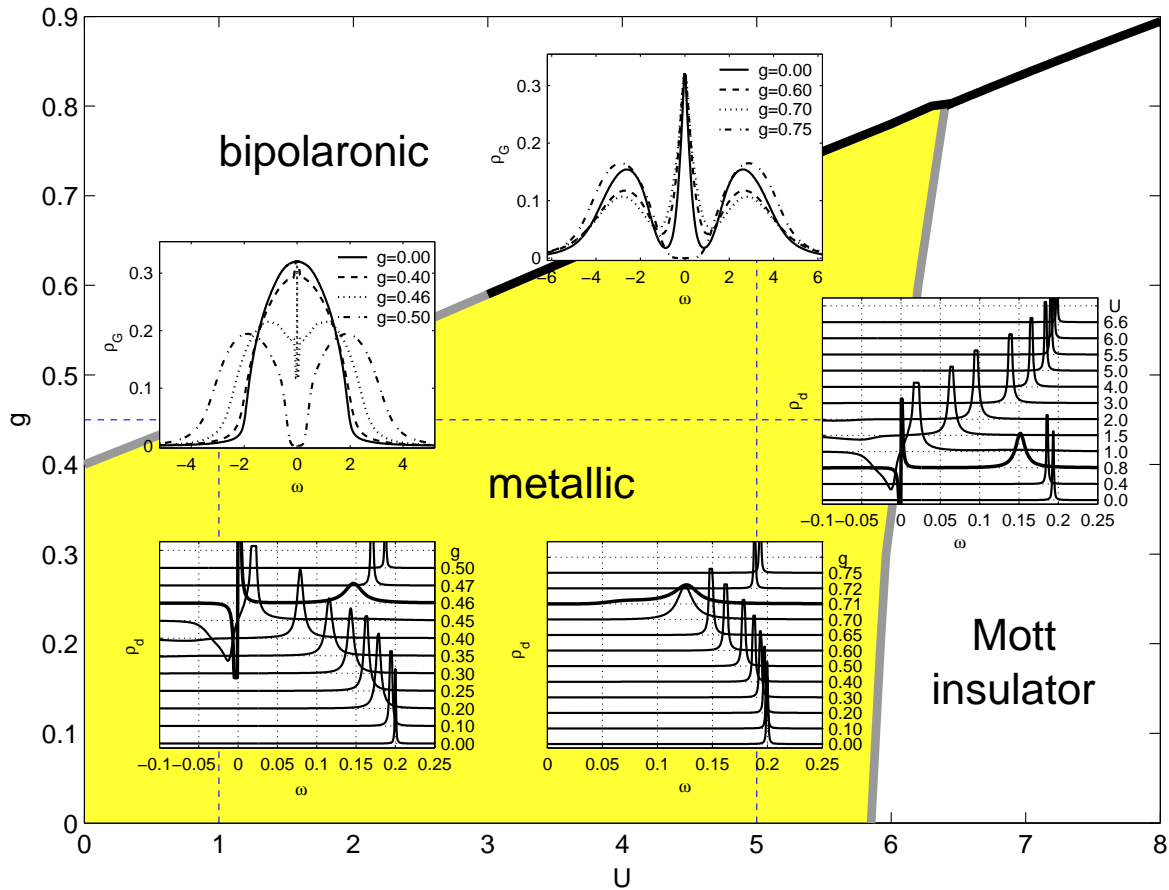}
    \caption{Phase diagram for the particle-hole symmetric Holstein-Hubbard
    model at zero temperature for a semielliptical band of width $W=4$ and a
    phonon frequency $\omega_0=0.2$.
    The insets show scans of the electron and phonon spectra, see text.}
\end{figure*}

All methods agree on the phase diagram shown in the figure.
It consists of a metallic region surrounded by two distinct gapped phases, a
Mott insulator for large $U$ and a bipolaronic phase, when the electron-phonon
coupling $g$ dominates.
The transition to the Mott insulating phase is always continuous,
as indicated by the grey boundary curve; the critical value of $U$ is largely
independent of $g$. In contrast, the transition to the bipolaronic phase is
first order for $U\gtrsim 3$, as indicated by the full black line, but
is also continuous for smaller values of $U$.

More of the physics of the interplay of the electron-phonon and
electron-electron interactions is revealed in the spectra of the dynamic
response functions. These were calculated using the DMFT-NRG \cite{MHB02}.
Electron and phonon spectra are shown within the figure along different
scans in the phase diagram.
The results for the corresponding dynamic spin and charge susceptibilities are
discussed in \cite{KMH04pre}.

The two upper insets show the electron spectra $\rho_G(\omega)$. The left-hand scan
for $U=1$ and $g=0.0, 0.4, 0.46, 0.5$ shows the continuous narrowing of the central
resonance with increasing $g$ and the subsequent opening of a gap at
$g_c\approx 0.47$.
Contrasting behaviour is seen in the upper right inset ($U=5, g=0.0, 0.6, 0.7,
0.75$) with an initial broadening of the central peak followed by its sudden
disappearance at a critical $g\approx 0.71$.

The two insets below the electron spectra show the behaviour of the
corresponding phonon spectra
 $\rho_d(\omega) = -\frac{1}{\pi}\textrm{Im} \biggreen{\bnod_i;\bdag_i}_\omega$.
For $U=1$ we see with increasing $g$ a complete softening of the phonon
peak as the transition to the bipolaronic phase is approached.
The increase of negative spectral weight for  $\omega<0$ indicates a 
growing number excited phonons.
At $g_c\approx 0.47$ we observe a two-peak structure whose low energy peak
vanishes and whose high-energy peak hardens back to $\omega_0$ after the
continuous transition.
For $U=5$ the softening of the phonons is much weaker due to suppression of
charge fluctuations. There is no two-peak structure at the transition and again
the peak hardens to $\omega_0$ in the gapped phase.

Similar features can also be seen in the scan for a fixed $g=0.45$ and
increasing $U$, shown in the right-hand inset.
The transition from the bipolaronic to the metallic phase is reflected by a
two-peak structure. Increasing $U$ further suppresses charge fluctuations and
hardens the completely softened phonon peak until the Mott transition is
approached and the peak arrives at the bare frequency $\omega_0$. \\[1ex]



\noindent
  {\bf Acknowledgements:}
  We wish to thank the EPSRC (Grant GR/S18571/01) for financial
  support. One of us (Y\=O) was supported by the
  Grant-in-Aid for Scientific Research from the Ministry of Education,
  Culture, Sports, Science and Technology.
  We also thank M. Aichhorn, R. Bulla, D. Edwards and  M. Potthoff for helpful
  discussions.

%
%
%
%

%
%
%
%


\end{document}